\pdfoutput=1

\documentclass[final, 10pt, conference]{IEEEtran}
\IEEEoverridecommandlockouts
\usepackage{cite}
\usepackage{amsmath,amssymb,amsfonts}
\usepackage{graphicx}
\usepackage{textcomp}
\usepackage[dvipsnames]{xcolor}
\usepackage{braket}
\usepackage{hyperref}
\hypersetup{
    colorlinks=true,
    linkcolor=BurntOrange,
    urlcolor=BrickRed,
    citecolor=Mulberry,
}
\usepackage{booktabs}

\begin{document}
\title{Exact and approximate simulation of large quantum circuits on a single GPU}

\author{
    \IEEEauthorblockN{
    Daniel Strano\IEEEauthorrefmark{1},
    Benn Bollay\IEEEauthorrefmark{4},
    Aryan Blaauw\IEEEauthorrefmark{4},
    Nathan Shammah\IEEEauthorrefmark{1},
    William J. Zeng\IEEEauthorrefmark{1}\IEEEauthorrefmark{3},
    Andrea Mari\IEEEauthorrefmark{2}\IEEEauthorrefmark{1}}
    \IEEEauthorblockA{\IEEEauthorrefmark{1}\textit{Unitary Fund}}
   \IEEEauthorblockA{\IEEEauthorrefmark{4}\textit{Independent researcher}}
    \IEEEauthorblockA{\IEEEauthorrefmark{3}\textit{Goldman, Sachs \& Co, New York, NY}}
    \IEEEauthorblockA{\IEEEauthorrefmark{2}\textit{Physics Division, School of Science and Technology, Universit\`a di Camerino, 62032 Camerino, Italy}}
}

\maketitle

\begin{abstract}
We benchmark the performances of Qrack, an open-source software library for the high-performance classical simulation of (gate-model) quantum computers. Qrack simulates, in the Schr\"odinger picture, the exact quantum state of $n$ qubits evolving under the application of a circuit composed of elementary quantum gates. Moreover, Qrack can also run approximate simulations in which a tunable reduction of the quantum state fidelity is traded for a significant reduction of the execution time and memory footprint.  
In this work, we give an overview of both simulation methods (exact and approximate), highlighting the main physics-based and software-based techniques. Moreover, we run computationally heavy benchmarks on a single GPU, executing large  quantum Fourier transform circuits and large random circuits. Compared with other classical simulators, we report competitive execution times for the exact simulation of Fourier transform circuits with up to 27 qubits. We also demonstrate the approximate simulation of all amplitudes of random circuits acting on 54 qubits with 7 layers at average fidelity higher than $4\%$, a task commonly considered hard  without super-computing resources.
\end{abstract}

\maketitle

\section{Introduction} 

The last decade was characterized by significant technological progress in quantum computing.
Several prototypes of quantum computers are currently available and are routinely used for research purposes and for proof-of-concept applications \cite{preskill2018quantum}.
It is not surprising that, at the same time, there has been a parallel progress in the classical simulation of quantum computers \cite{xu2023herculean, Pednault_2017_Arxiv, Strano_vm6502q_qrack, Fatima_2021_HPCA, villalonga2019flexible, huang2021efficient, gray2021hyper, vincent2022jet, kelly2018simulating, roberts2019tensornetwork}.

Developing powerful and efficient classical simulators of quantum computers is important for several reasons. A first reason is to numerically test quantum algorithms applied to a limited number of qubits, without the need of using expensive quantum hardware. A second reason is to calibrate and validate noisy quantum computers, since this typically requires a comparison between the noisy results of the real device against the ideal results of an exact classical simulation. A third reason is to run algorithms where quantum processors and simulated processors cooperate in a hybrid computation.  A further motivation for classically simulating quantum computers is to provide an empirical baseline of performances (e.g. \cite{huang2020classical, pan2021simulating, pan2021solving}) which, at least in principle, should be overcome by real devices in order to demonstrate any claim of {\it quantum advantage} \cite{bravyi2018quantum, huang2021information, daley2022practical, huang2022quantum, deshpande2022quantum, zhong2020quantum} or {\it quantum supremacy} \cite{harrow2017quantum,aaronson2016complexity, arute2019quantum}.
Beyond the mentioned applications, one should not underestimate the importance of developing new  quantum-inspired computational paradigms from a fully classical computer science perspective. In fact, as we also show in this work, quantum-inspired algorithms might match or exceed the performances of standard classical algorithms, especially if the transparent  parallel nature of quantum dynamics is exploited for GPU acceleration or for HPC execution.  

Here we focus on a specific library for the simulation of gate-model quantum computers: {\it Qrack} \cite{Strano_vm6502q_qrack}.
Qrack is an open-source framework founded in 2017, undergoing continuous development up to the present days. It is designed to serve, with high performances, all scales of simulation: from single consumer CPUs to an arbitrarily high number of clustered GPUs.

The main contribution of this work is to present the optimization and approximation techniques underlying the Qrack software library and to validate its simulation performance against computationally intensive benchmark problems. 

Specifically, we run exact simulations of quantum Fourier transform (QFT) circuits up to $27$ qubits and we run approximate simulations of random circuits at 54 qubits, up to 10 layers, with average fidelity estimated at $\approx 4\%$ at 7 circuit layers, on a single A100 GPU in very short execution time. Differently from previous works in the literature, we do not use advanced super-computers or expensive multi-GPU cloud services for running our simulations. All QFT results presented in this work are obtained with a single laptop, and all approximate results are run on a single GPU (NVIDIA A100). These can thus be reproduced with relatively little cost by the scientific community. High performance simulation using limited computational power with Qrack is realized through several ingredients that will be explained in the main sections of this work: circuit simplification techniques, a Schmidt decomposition optimization, a tunable Schmidt-decomposition-rounding-parameter (SDRP) approximation, and other synergistic optimizations that include ``hybrid'' stabilizer/ket simulation and well-rounded use of proven and novel HPC software engineering techniques with adherence to disciplinary best-practices (see Table \ref{table:qrack-methods-main}). 

This work is organized in two main sections. In Sec.\ \ref{sec:exact} we restrict the analysis to the exact simulation of quantum circuits and we benchmark the execution time of quantum Fourier transform circuits. In Sec.\ \ref{sec:approximate}, we instead focus on the approximate simulation of quantum circuits, we describe the Schmidt decomposition rounding parameter (SDRP) technique and we approximately simulate large-scale random circuits.

\section{Exact simulation} \label{sec:exact}

The first task that we consider is the exact simulation of a quantum circuit in the Schr\"odinger picture.
Let $|\psi_C\rangle = C |\psi_0\rangle$ be the {\it ket} quantum state obtained by the application of a quantum circuit $C$ to some given initial state $|\psi_0\rangle$ of $n$ qubits. Our goal is to compute $|\psi_C\rangle$ given $|\psi_0\rangle$ and a description of $C$ in terms of a sequence of local gates acting on $k$ qubits (with $k$ typically equal to 1, 2 or 3).

The Qrack simulation approach is based on the following four general principles.

\subsection{Keeping ket states as factorized as possible} \label{subsec:factorize}
During the classical simulation steps that are necessary to compute the state evolution, Qrack keeps the state  representation {\it as factorized as possible} to increase the simulation efficiency \cite{wecker2014liqui}.

A generic ket state $|\psi\rangle$ of $n$ qubits is characterized by $\mathcal O(2^n)$ complex amplitudes. However, if the state $|\psi\rangle$ can be factorized as the tensor product of $m$ local states 
\begin{equation}
    |\psi\rangle=|\psi_{S_1}\rangle \otimes |\psi_{S_2}\rangle \dots |\psi_{S_m}\rangle, 
\end{equation}
where $S_1, \dots S_m$ represent disjoint subsets of the $n$ qubits, the number of complex amplitudes that are necessary to represent $|\psi\rangle$ can be significantly  reduced.
For example, in the extreme limit $m=n$, i.e. when the qubits are not entangled, $\mathcal O(n)$ complex amplitudes are sufficient. A more realistic example, which is quite relevant for Qrack, is the case where all qubits are highly entangled with the exception of a single qubit $q$ which can be fully factorized, i.e. $|\psi\rangle = |\psi_{q} \rangle \otimes |\psi'\rangle$. In this case, which Qrack is able to detect during the simulation process as shown in Sec.\ \ref{subsec:sdrp}, the representation cost is given by $\mathcal O(2^{n-1})$ complex amplitudes, corresponding to halving the cost associated to a fully entangled state.

\subsection{Any ``unobservable" circuit optimization is allowed} \label{subsec:unobservable}
Qrack modifies, simplifies  and sometimes completely removes some gates of the simulated circuit whenever this has {\it unobservable} consequences.

More precisely, we say that a circuit transformation $C \rightarrow C'$ is  unobservable with respect to the specific input state $|\psi_0\rangle$, if 
\begin{equation}
    \left| \langle \psi_0 | C'^\dagger C|\psi_0\rangle \right|^2=1 .
\end{equation}
In other words, replacing the original circuit $C$ with a simplified circuit $C'$ is always allowed as long as the results of the computation are unaffected.

Note that $C$ and $C'$ may correspond to different unitary operations but, as long as they are equivalent when applied to the specific input state $|\psi_0\rangle$, we can safely change $C$ into $C'$.
For example, if the control qubit of a {\rm CNOT} gate is in the state  $|0\rangle$, the CNOT gate can be removed. Similarly, if the control qubit is in the state $|1\rangle$ a CNOT gate can be replaced by a local bit-flip of the  target qubit.

From this simple example it is clear that, in order to determine if a gate optimization is unobservable, it is crucial to have direct access to the ket state at each step of the simulation (Schr\"odinger picture).
This is a specific design feature which is typically not present in standard  tensor-networks simulators in which the ket state is not explicitly accessible, but must be computed with a non-negligible cost. 

\subsection{Hybridizing ket and stabilizer representations} \label{subsec:stabilizer}

It is well known that Clifford circuits can be classically simulated with a polynomial cost \cite{Aaronson_2004_PRA}. This fact is exploited by Qrack which applies an hybridized stabilizer/ket simulation approach. In practice, one can  detect which parts of the computation are more convenient to simulate via stabilizer tableaus and which parts are more convenient to simulate via ket states. Beyond the obvious application in simulating fully Clifford circuits, this hybrid approach is particularly convenient also in circuits with deep  Clifford or (near-Clifford) preambles.

Qrack chooses ``transparently'' (i.e. by default) between stabilizer and ket simulation. The library introspects whether calculations are being carried out as stabilizer, and it enacts Clifford entangling gates with priority if they would be carried out in stabilizer formalism, otherwise attempting to buffer them to the benefit of state factorization. Ultimately, the stabilizer-hybridization layer of Qrack relies on fallback from stabilizer to ket representation in all cases which cannot be reduced to Clifford (as through unobservable circuit optimizations). The stabilizer hybridization layer also assigns a general single-qubit unitary gate buffer respectively to each qubit, which enables further optimization by simple commutation, to maintain underlying Clifford representation. One key factor contributing to the efficiency of Qrack's stabilizer-hybridization layer is aggressive state factorization as a higher and external priority, including for stabilizer subsystems.

\subsection{Optimizing computational resources} \label{subsec:computational}
In addition to the previous physics-based optimization  techniques, Qrack makes use of software-based techniques to achieve high performances with minimal computational resources.
Qrack is written in ``pure language'' C++11 standard with only optional OpenCL (or CUDA) and Boost library dependencies and licensed header reuse. Because of its minimal number of  dependencies, a ``full-feature'' Linux build of the library might require about 16 MB of disk footprint, or about 4 MB when compressed, and this can be further reduced for custom builds. The comparatively extreme compactness of the compiled library likely also benefits use and management of CPU cache.
Additionally, CPU/GPU hybridization is supported through the \texttt{QHybrid} class (see Table \ref{table:qrack-methods-main}) allowing for an optimal distribution of computational resources, such that CPU is used for small circuits and GPU is preferred for large circuits (see e.g. Fig.~\ref{fig:cpu-gpu} in Appendix \ref{appendix-benchmarks}). 
Finally, Qrack also supports single instruction multiple data (SIMD) intrinsics (see Table \ref{table:qrack-methods-main}) for obtaining the advantages of data vectorization at the CPU level.

%\iffalse
\begin{table*}[!htb]
\setlength{\tabcolsep}{2.5em} % for the horizontal padding
{\renewcommand{\arraystretch}{1.5}
    \caption{Optimization methods employed by Qrack with associated pictorial representation, section and reference.}
    \label{table:qrack-methods-main}
    \begin{tabular}{  p{5.4cm} p{4.4cm}  p{1cm} p{1cm}}
        \toprule
        \\
\textbf{Technique}      
&
\textbf{Diagram}
&
\textbf{Section}
& \textbf{Ref.}\\
\\
\toprule
Ket simulation is proactively and reactively Schmidt decomposed.
&
\raisebox{-10mm}{\includegraphics[width=23mm]{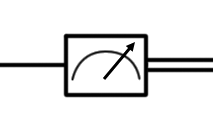}}
& 
Sec.~\ref{subsec:factorize}
&
\cite{Pednault_2017_Arxiv}
\\\hline
SWAP gate is constant complexity (i.e. via qubit label swap).
&
\raisebox{-10mm}{\includegraphics[width=33mm]{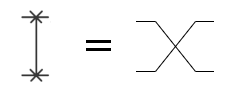}}
&
Sec.~\ref{subsec:unobservable}
&
Qrack (2018)
 \\\hline
\texttt{QUnit} eliminates Z-basis eigenstate controls.
&
\raisebox{-6mm}{\includegraphics[width=33mm]{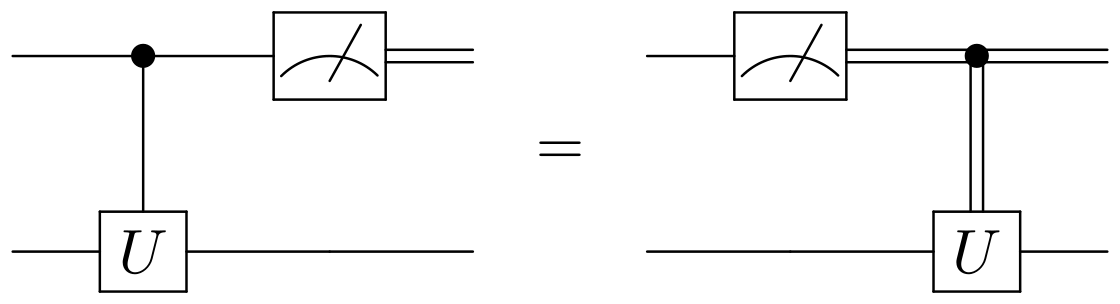}}
&
Sec.~\ref{subsec:unobservable}
&
Qrack (2018)
 \\\hline 
Controlled phase and “inversion” (X-like) gates are buffered and commuted around H gate.
&
\raisebox{-7mm}{\includegraphics[width=38mm]{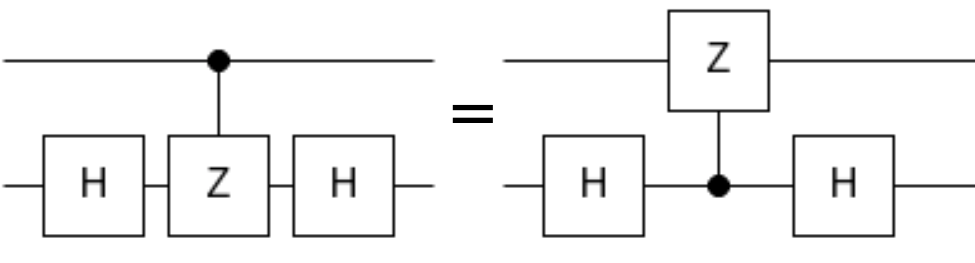}}

& 
Sec.~\ref{subsec:unobservable}
&
Qrack 2019
 \\\hline
Simulation is “hybridized” with transparent switch-off between stabilizer and ket “shards”
&
\raisebox{-10mm}{\includegraphics[width=28mm]{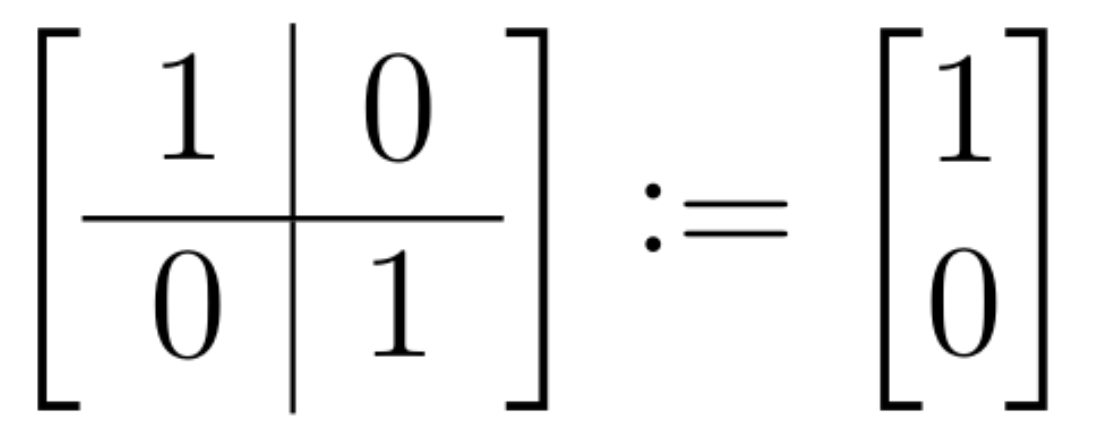}}
&
Sec.~\ref{subsec:stabilizer}
&
\hspace{-0.1em}\cite{Aaronson_2004_PRA}, 
 Qrack v.5.4.0 (2021)
  \\\hline
Coalesced parallel X/Y/Z gates are executed with single traversal.
&
\raisebox{-6mm}{\includegraphics[width=33mm]{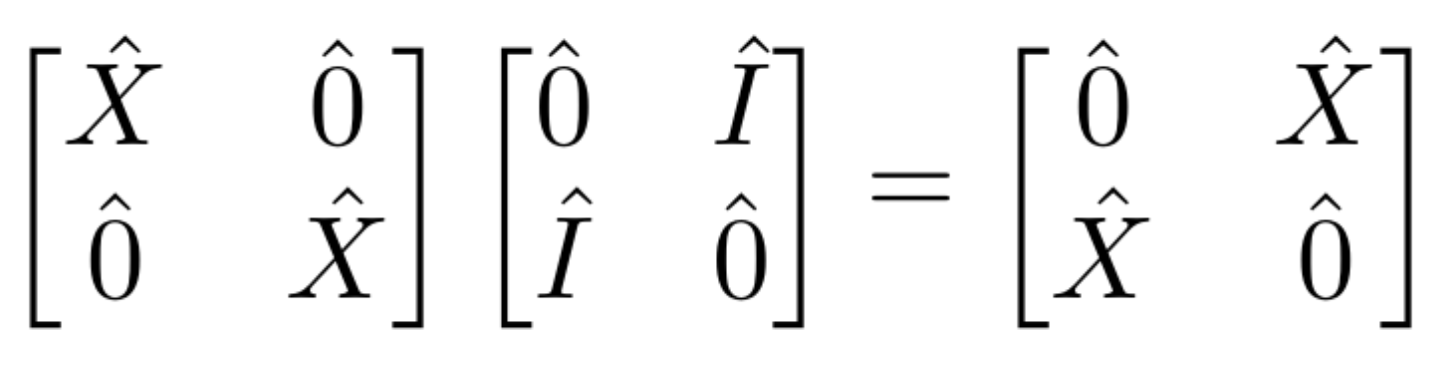}}
&
Sec.~\ref{subsec:computational}
&
\cite{Fatima_2021_HPCA}
 \\\hline
Ket simulation ``hybridizes'' CPU/GPU methods.
&
\hspace{6mm}\raisebox{-24mm}{\includegraphics[width=18mm]{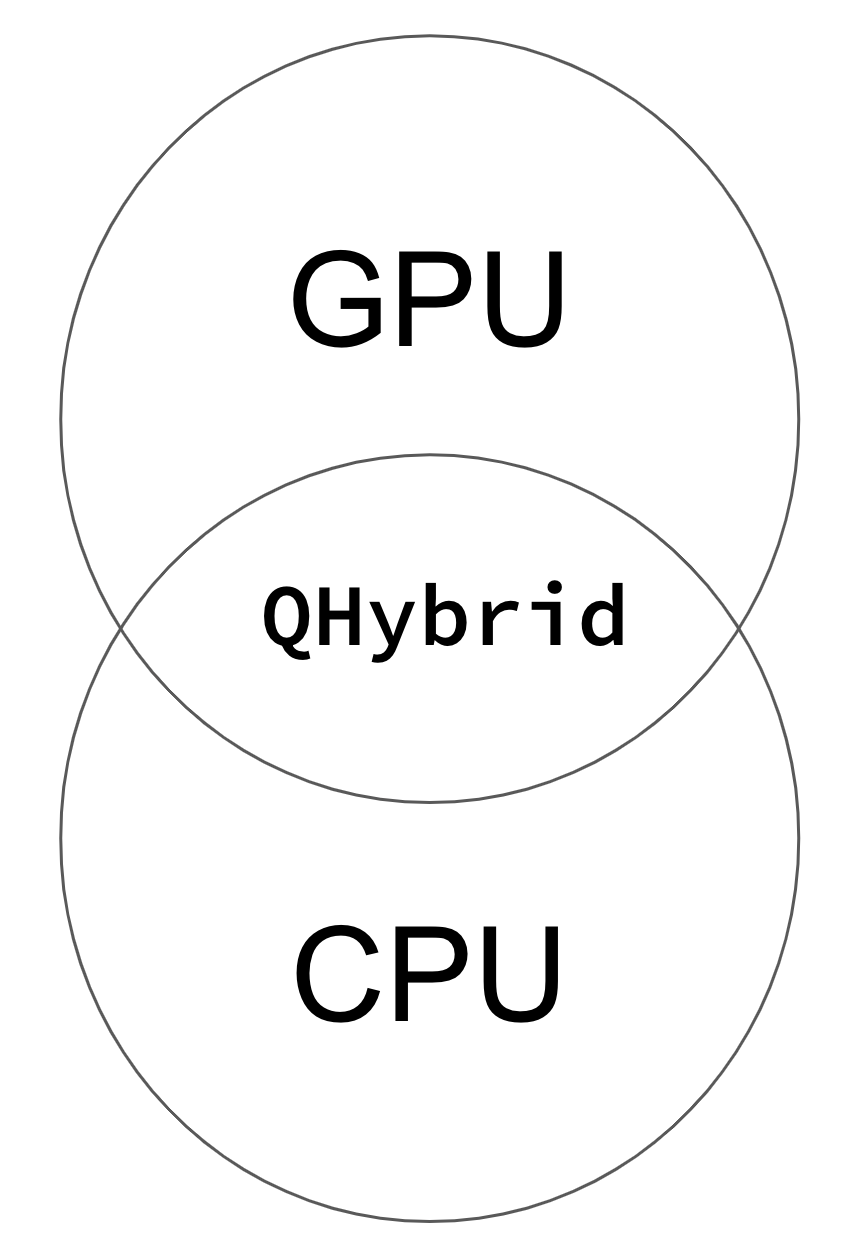}}
&
Sec.~\ref{subsec:computational}
&
Qrack 2020
 \\\hline
SIMD (SSE and AVX) instruction sets of x86 and x86\_64 processors improve gate-application throughput.
& 
\hspace{5mm}\raisebox{-18mm}{\includegraphics[width=22mm]{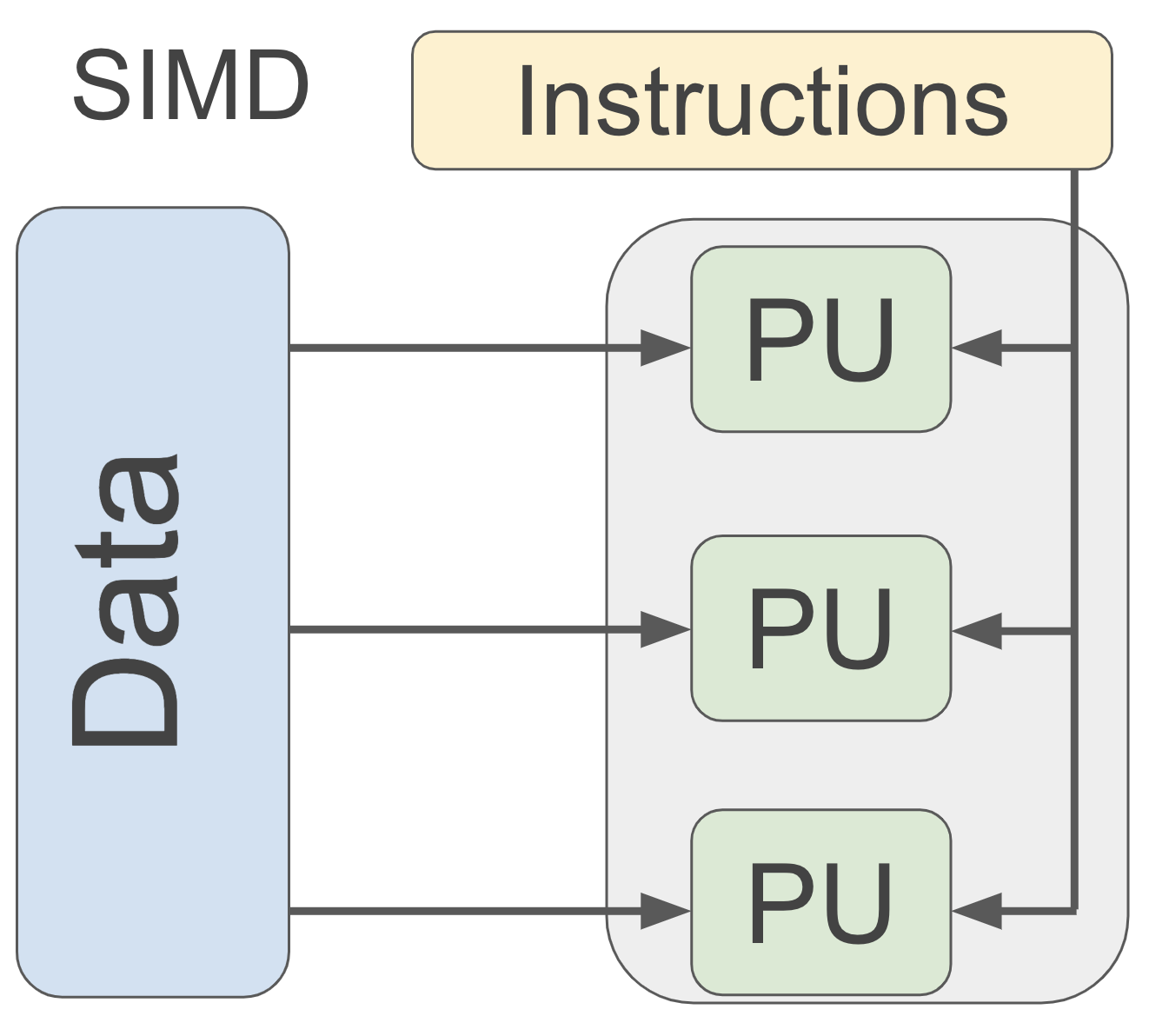}}

&
Sec.~\ref{subsec:computational}
&
SSIS x86/ x86\_64 CPUs
 \\\hline
 
        \toprule
    \end{tabular}
}
\end{table*}

%\fi

\begin{figure*}[!ht]
\begin{center}
  \includegraphics[width= 0.49 \linewidth]{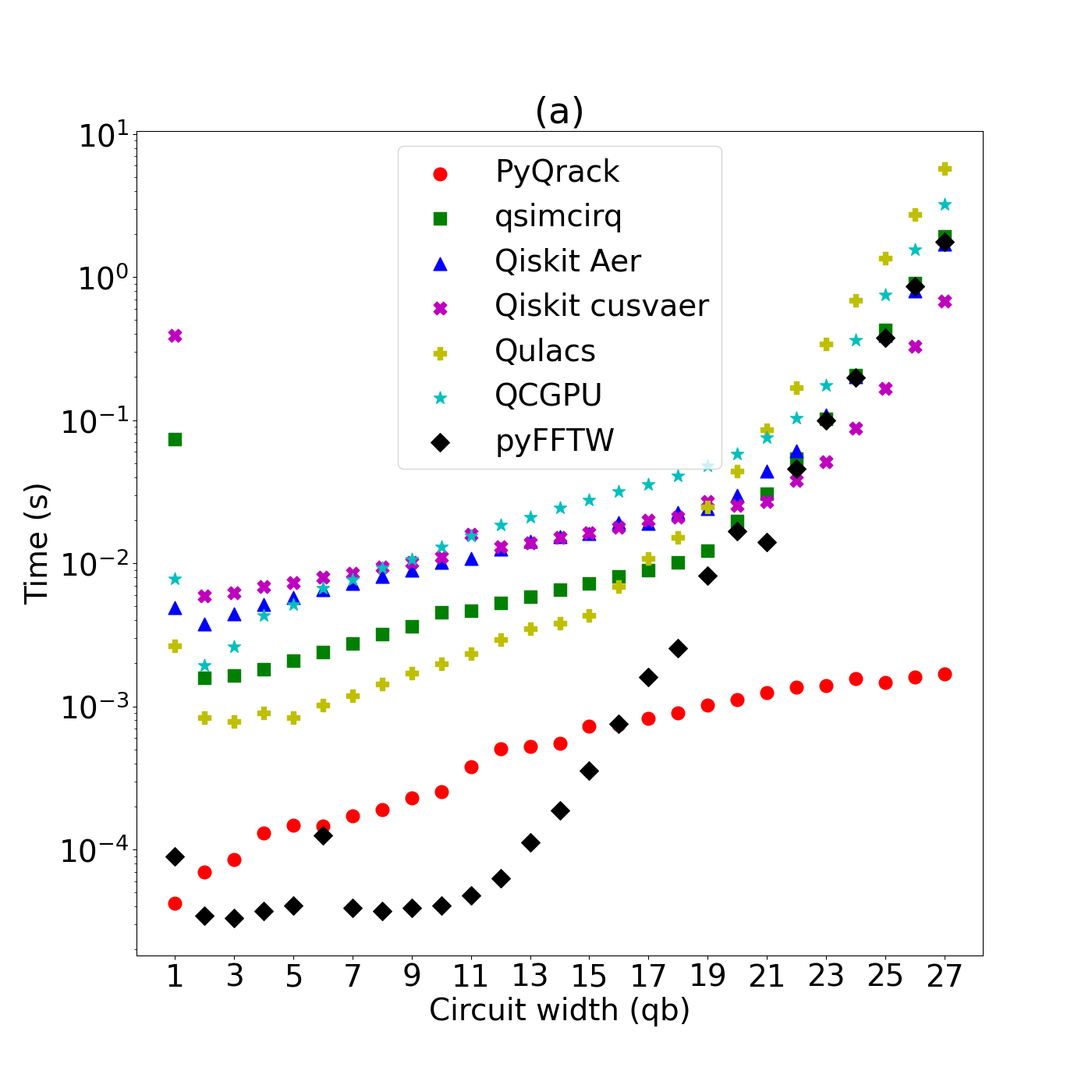}
    \includegraphics[width= 0.49 \linewidth]{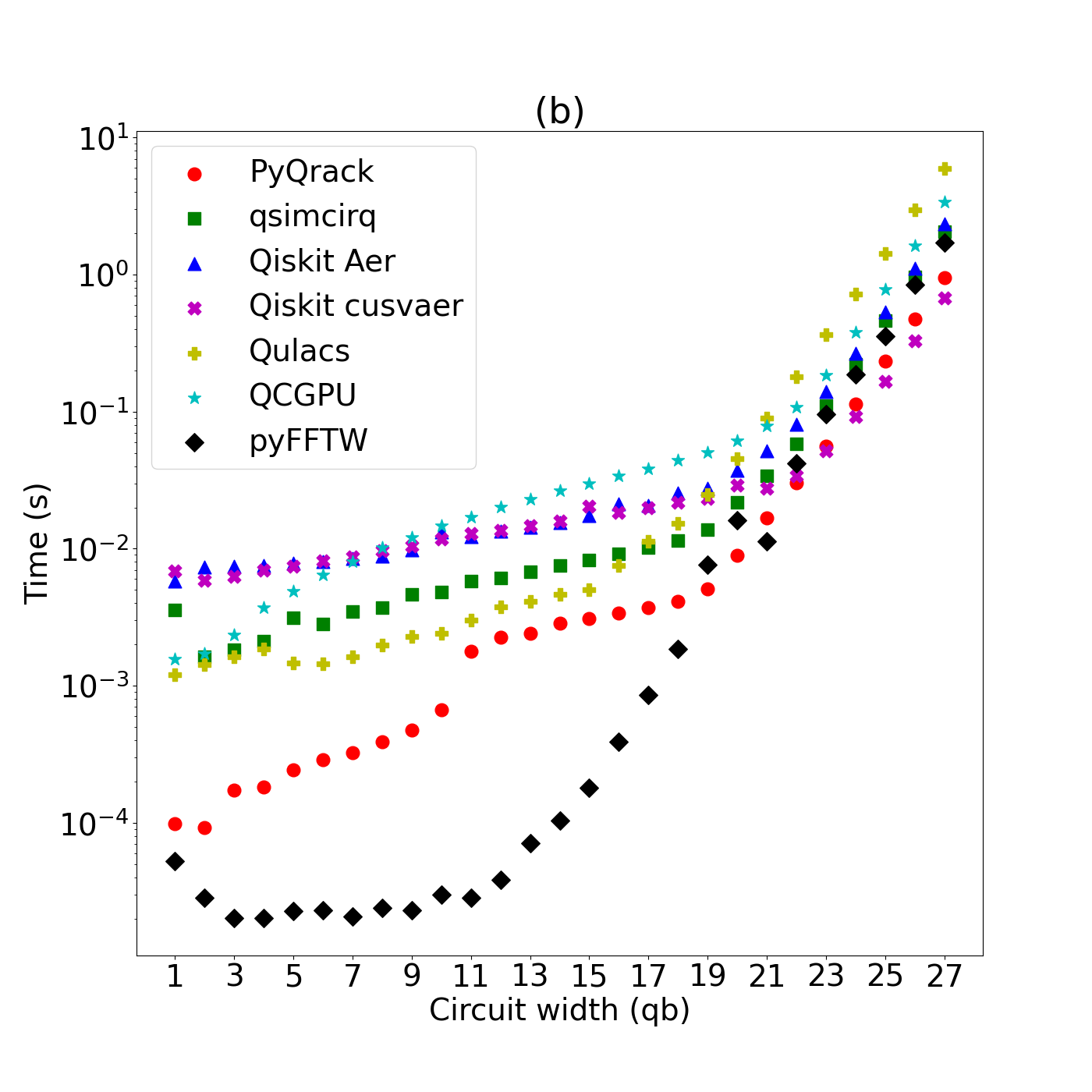}
  \caption{(a) Wall-clock execution time of quantum Fourier transform (QFT) circuits applied to the initial state $|0\dots 0\rangle$ and executed with different classical simulators. (b) Repetition of the same  benchmark but using the GHZ state defined in Eq.\ \eqref{eq:ghz} as initial condition. Because of its large amount of entanglement, the GHZ state can be considered as the worst-case initialization scenario for Qrack. For both plots, all candidates were executed on the same Alienware m17 laptop, with Alienware BIOS version 1.16.2, (BIOS overclocking features set to off/default,) Ubuntu 22.04 LTS, Linux kernel version 5.19.0-35-generic, one ``Intel(R) Core(TM) i9-10980HK CPU @ 2.40GHz,'' one ``NVIDIA GeForce RTX 3080 Laptop GPU,'' and 32 GB of SK hynix 3200 MT/s DDR4 in 2x16 GB row configuration, collected on ISO date 2023-03-24. Candidate release versions were PyQrack 1.4.2 \cite{Strano_vm6502q_qrack}, qsimcirq 0.12.1 \cite{Cirq} (in NVIDIA appliance Docker image), Qiskit Aer 0.12.0 \cite{Qiskit}, Qiskit Aer 0.11.0 (for cusvaer \cite{cuQuantum} in NVIDIA appliance Docker container), Qulacs 0.5.7.dev78+gc3b28f13 \cite{suzuki2021qulacs}, QCGPU 0.1.1 \cite{kelly2018simulating}, and pyFFTW 0.13.1. \cite{frigo2005design} }
  \label{fig:qft}
\end{center}
\end{figure*}

\subsection{Simulation benchmark: quantum Fourier transform}

To test the efficiency of Qrack with respect to the task of exactly simulating quantum circuits, we consider numerical experiments based on the quantum Fourier transform (QFT).

QFT circuits are often used for validating real or simulated quantum computers. Here we use them to benchmark Qrack against similar GPU-based classical simulators \cite{kelly2018simulating, Qiskit, Cirq, suzuki2021qulacs, cuQuantum} and against a fully  classical algorithm \cite{frigo2005design} for computing discrete Fourier transforms.

Let $|\psi_0\rangle$ be an arbitrary initial state of $n$ qubits. We can represent it in the computational basis as $|\psi_0\rangle = \sum_j x_j |b_j\rangle$ where $\{|b_j\rangle\}$ are the basis elements labelled with bitstrings $b_j$ corresponding to the binary representation of the integers $j=0, 1, 2, \dots N$,  with $N=2^n$.
The quantum Fourier transform is the unitary operation $C_{\rm QFT}$ which acts as follows:

\begin{equation}\label{eq:qft}
    |\psi_0\rangle = \sum_{j} x_j |b_j\rangle  \rightarrow C_{\rm QFT}|\psi_0\rangle = \sum_{j} y_j |b_j\rangle,
\end{equation}
where the complex  amplitudes $\{y_j\}$ are (up to a different sign convention) the classical discrete Fourier transform (DFT) \cite{brigham1967fast} of the input amplitudes $\{x_j\}$:
\begin{equation}\label{eq:dft}
    y_j = \frac{1}{\sqrt{N}} \sum_{j=0}^{N-1} x_k \omega^{j k}, \quad \omega = e^{\frac{i 2 \pi }{N}}.
\end{equation}

Importantly, it is known how to decompose the unitary $C_{\rm QFT}$ as a circuit of two elementary gates: the Hadamard gate H and the controlled version of the single qubit phase gate $R_Z(\theta)$. So, given its decomposition as a quantum circuit, we can measure the wall-clock time for simulating $C_{\rm QFT}$ with different simulators obtaining simple and reproducible benchmarks.\footnote{Our choices of convention for the final state of the (inverse) QFT run by all simulators match the DFT up to normalization.}

Moreover, since the QFT circuit $C_{\rm QFT}$ acts on state amplitudes according to the classical DFT defined in Eq.\ \eqref{eq:dft}, we also add a fully classical DFT library (pyFFTW \cite{frigo2005design}) within the set of the benchmarked simulators. We stress that, technically, pyFFTW is not an actual simulator since it is not able to simulate quantum circuits. On the other hand, pyFFTW is one of the fastest classical algorithms for computing Eq.\ \eqref{eq:dft} and therefore it provides a useful benchmark for the performances of quantum simulators acting on QFT circuits.

The results are reported in Figures \ref{fig:qft}a and \ref{fig:qft}b.

In Figure \ref{fig:qft}a, the initial state is $|0 \dots 0 \rangle$, i.e., all qubits are in the configuration $|0\rangle$, the standard initial state in most quantum computing algorithms. In this case, Qrack is able to exploit the state factorization of the initial state to execute the QFT circuits several order of magnitude faster than other simulators and, within the limits of our numerical analysis, with a better asymptotic scaling.
Remarkably, in this case Qrack clearly outperforms even the classical DFT algorithm pyFFTW for $n>16$. This is an interesting consequence of the fact that Qrack is highly optimized to simulate factorized states.

In Figure \ref{fig:qft}b we repeat the same numerical experiment with an initial state which is more difficult to handle by circuit simulators  because of its large amount of entanglement. Specifically we use the GHZ entangled state:
\begin{equation}\label{eq:ghz}
|\psi_{\rm GHZ}\rangle = \frac{1}{\sqrt{2}}\left(|0\dots 0\rangle + |1 \dots 1\rangle \right).
\end{equation}
Also in this case, Qrack nearly outperforms all the considered classical simulators of quantum circuits, with the exception of a small advantage (less than a factor of 2) by Qiskit cusvaer (from the cuQuantum appliance for systems with NVIDIA GPUs) at high-widths ($n \geq 23$). When compared to the classical DFT library pyFFTW, Qrack is slower for circuits having a small number of qubits  ($n \le 20$) and is slightly faster for larger circuits ($n>20$).

It is worth noting that only the QCGPU and Qrack simulators can run on non-NVIDIA GPUs, as these two libraries use the OpenCL API for general purpose GPU and accelerator programming, while all the other simulators shown in Fig.\ \ref{fig:qft} are based upon the proprietary CUDA API which can be used with NVIDIA GPUs only. Qrack alone also optionally supports CUDA as an alternative to OpenCL. pyFFTW also runs on virtually any system with a CPU, since it does not support GPU acceleration at all.  Additional information about the Qrack  simulation time with and without using GPU acceleration, is given in Fig.~\ref{fig:cpu-gpu} of Appendix \ref{appendix-benchmarks}.

\section{Approximate simulation} \label{sec:approximate}

\begin{figure*}[!htb]
  \includegraphics[width=\linewidth]{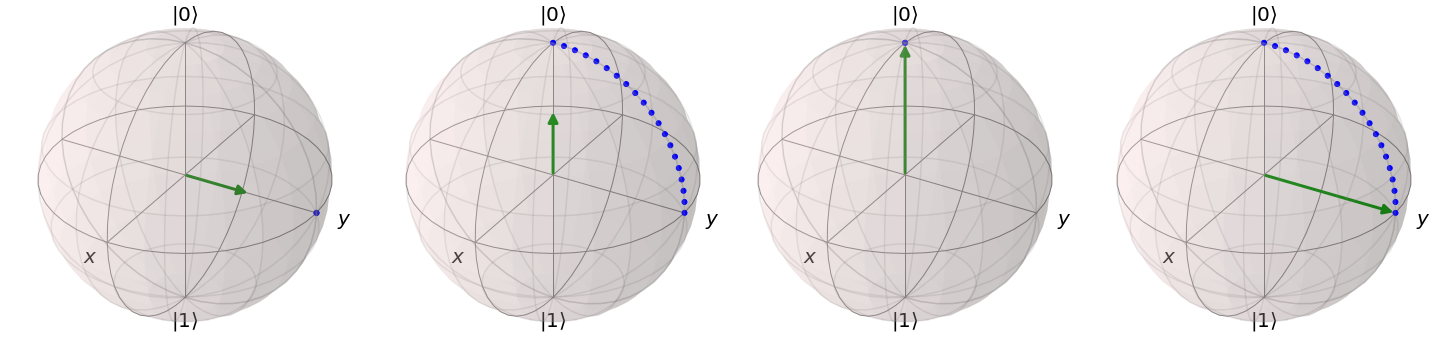}
  \caption{Pictorial representation of the SDRP approximation technique. We represent the reduced state of a qubit as a vector in the Bloch sphere (first image) and we check if its length is longer than $1-p$, for threshold parameter $p$. If so, we rotate the state along the direction of the $|0\rangle$ pole (second image). We then post-select the measurement of the $|0\rangle$ state, extending the length of the Bloch vector to 1 (third image). We finally reverse the original rotation, such that the Bloch vector points along the original axis (fourth image).}
  \label{fig:sdrp-bloch-sphere}
\end{figure*}

In Sec.\ \ref{sec:exact}, we focused on the {\it exact} simulation of quantum circuits.
In this section instead, we consider the {\it approximate} simulation of quantum circuits.

Limited cases of exact high-width simulation are already possible in Qrack due to its factorization of subsystems and stabilizer tableau capabilities. However, many realistic circuits still require a peak memory footprint which is close to the footprint of a brute-force simulation. For large circuits ($n>30$), this limits Qrack's exact simulation capabilities. Compared to state-of-the-art in tensor network simulations, this is a major limitation of Qrack exact simulation methods. This motivated the development of an approximate simulation method designed to trade minimal fidelity loss for maximum reduction of memory and time complexity.

\subsection{Schmidt decomposition rounding parameter approximation} \label{subsec:sdrp}

If we isolate a single qubit $q$ from the associated  complementary set $\bar q$ of $n-1$ qubits, we can always express the full ket state of the system with the following Schmidt decomposition: 
\begin{equation}\label{eq:schmidt}
|\psi\rangle = (1 - \sqrt{\epsilon}) |\varphi \rangle_{q} |\psi \rangle_{\bar q} + \sqrt{\epsilon} |\varphi^{\perp}\rangle_{q} |\psi^{\perp}\rangle_{\bar q},
\end{equation}
where $|\varphi \rangle |_{q}$ is a quantum state of the qubit and $|\varphi^\perp \rangle |_{q}$  is its {\it unique} orthogonal state. Similarly, $|\psi \rangle_{\bar q}$ is a quantum state of all the other qubits $\bar q$ and $|\psi^\perp \rangle_{\bar q}$ is an orthogonal state.

Without loss of generality, we can assume $\epsilon \in [0, 0.5]$, such that $\epsilon$ can be used to quantify the amount of entanglement between the qubit $q$ and the rest of the system $\bar q$.  If $\epsilon$ is large, the state is highly entangled; if $\epsilon$ is small, the state is weakly entangled. If $\epsilon=0$ (up to machine precision) the qubit is fully separable and an exact simulation with a factorized ket representation is possible as discussed in Sec.\ \ref{sec:exact}.  In this section instead we are interested in approximating the state for values of $\epsilon$ which are small but nonzero.

More precisely, we define the Schmidt decomposition rounding parameter (SDRP) approximation, with threshold parameter $p \in [0, 1]$, as the following non-unitary operation:
\begin{align}\label{eq:sdrp}
&|\psi\rangle \rightarrow |\varphi \rangle_{q} |\psi \rangle_{\bar q}, \quad \text{if } \epsilon \le p/2,  \\
&|\psi\rangle \rightarrow |\psi\rangle ,  \qquad \quad \text{ if } \epsilon > p / 2. \nonumber
\end{align}
In other words, if the state is weakly entangled, the SDRP approximation projects the state on the closest factorized state corresponding to the dominant term in the Schmidt decomposition. Else, no approximation is applied.

In practice, the way in which Qrack implements the SDRP approximation defined in Eq.\ \eqref{eq:sdrp} is through four geometrical transformations representable in  the Bloch sphere (see Fig.\ \ref{fig:sdrp-bloch-sphere}):

\begin{enumerate}
\item Since we have access to the ket state $|\psi\rangle$, we can easily compute the local expectation values of the $X_q$, $Y_q$ and $Z_q$ Pauli operators associated to the qubit $q$. Therefore, we have full knowledge of the reduced (mixed) state of $q$:
\begin{equation}
\rho_q = \frac{1}{2}[I_q + r_x X_q + r_y Y_q + r_z Z_q]
\end{equation}
which is completely characterized by the 3D Bloch-sphere vector 
\begin{equation}
{\bf r} =[r_x, r_y, r_z] = [\langle X_q \rangle, \langle Y_q \rangle, \langle Z_q \rangle ].
\end{equation}
We remark that the Schr\"odinger-picture simulation approach is crucial for quickly computing $\bf r$.

\item Given the Bloch vector $\bf r$, we deduce $\epsilon$ from its modulus, since it is easy to show that $\epsilon = (1 - |{\bf r}|)/2$. If $\epsilon > p/2$, no approximation is applied. If $\epsilon \le p/2$, a local unitary is applied to rotate $\bf r$ along the direction of the $|0\rangle$ pole.
\item In this new reference frame, the projection defined in Eq.\ \eqref{eq:sdrp} can be implemented as a simple measurement in the computational basis, post-selected on the $|0\rangle$ outcome, which yields the normalized state $|0\rangle_q |\psi\rangle_q$. This measurement-like step is a fundamental capability which must be obviously present in any classical simulator, including Qrack.
\item The state of the qubit is finally rotated back along the direction of the original Bloch vector $\bf r$, obtaining the desired final state $|\varphi \rangle_q  |\psi\rangle_q$.
\end{enumerate}

Note that, from an abstract point of view, this technique is similar to the core idea of matrix product states (MPS) \cite{schollwock2011density, vidal2003efficient, perez2006matrix}, but it is focused on the particular case in which one of the subsystems in the Schmidt decomposition is a single-qubit. (Qrack has also generalized the technique analogously to at least 2-qubit subsystems, but this does not factor into the particular benchmarks presented.) Within the MPS formalism, a common way to recover efficiency of approximate simulation is to represent states as tensors, perform a singular value decomposition, and discard principle components with small Schmidt coefficients (see e.g.\ \cite{fishman2022itensor} for a recent software  implementation). The peculiar aspect of the SDRP technique presented in this work is the possibility of applying the same type of Schmidt projection used in MPS {\it without} representing states as MPS, but by performing instead a task which is elementary for any classical simulator: rotating and measuring a single qubit in the computational basis. 

\subsection{Estimating the simulation fidelity} \label{fidelity_estimation}
Let us denote with $| \psi_j^{(in)} \rangle$ and $| \psi_j^{(out)} \rangle$ the simulated ket states evaluated right  before and right after the $j$th SDRP projection (with $\epsilon_j \le p$) applied during the   circuit simulation.
It is easy to check that, for  each individual approximation, the fidelity of the output state with respect to the input given by:
\begin{equation}
 |\langle \psi_j^{(out)}| \psi_j^{(in)}  \rangle|^2 = 1 - \epsilon_j.
\end{equation}
This fidelity reduction is due to neglecting the $\sqrt{\epsilon_j} |\varphi_j^{\perp}\rangle_q |\psi_j^{\perp}\rangle_q$ branch of Eq.\ \eqref{eq:schmidt}.
If all the neglected branches associated to multiple SDRP approximations remained orthogonal to the preserved branch along the full simulation, the fidelity of the final approximated state with respect to the ideal exact state would be
\begin{equation}\label{eq:fid_model}
\mathcal F = \prod_j (1 - \epsilon_j).
\end{equation}
In practice however, the Schmidt branches associated to different SDRP approsimations can have a small overlap with the simulated state, such that Eq.\ \eqref{eq:fid_model} is not an exact formula but, nonetheless,  is a very good estimator of the actual fidelity. 

To validate the model introduced in Eq. \eqref{eq:fid_model}, we compared it with the exact fidelity which, by definition, can be computed from the exact simulation of the final state:
\begin{equation}\label{eq:fid_exact}
\mathcal F_{\rm exact} =  |\langle \psi_{\rm approx} | \psi_{\rm exact} \rangle|^2.
\end{equation}
For random circuits of limited size, we find a very good agreement between the two quantities. (See Appendix \ref{appendix-validation} for more details on the statistical validation of the fidelity estimation model). This fact allows us to efficiently estimate the fidelity of large-scale simulations, without the need of computing \eqref{eq:fid_exact} which instead would require an exact simulation with huge computational resources.

\subsection{Simulation benchmark: random circuits}\label{subsec:circuits}

\begin{figure}[!ht]
\begin{center}
  \includegraphics[width= 1.1 \linewidth]{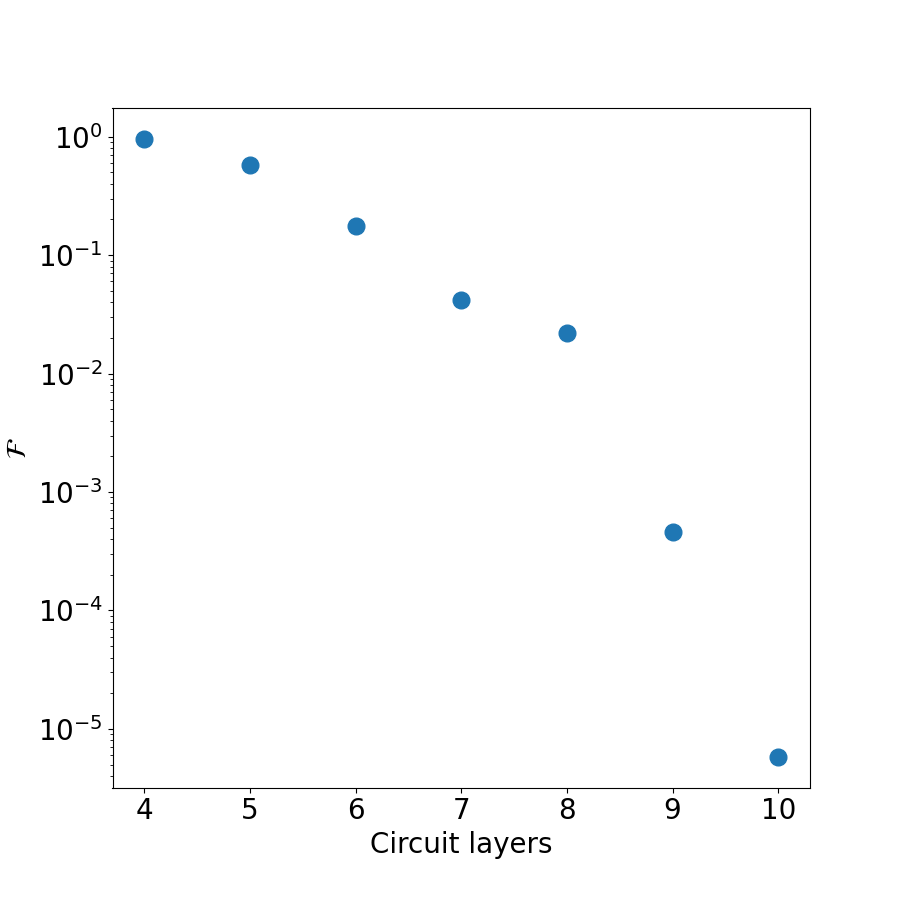}
  \caption{Achievable fidelity for the simulation of random circuits acting on $54$ qubits with $d$ layers on a single GPU. The fidelity is estimated by  the empirically-validated model discussed in Sec. \ref{fidelity_estimation} averaged over 100 random circuits for each data point. The cloud-compute virtual machine was a Paperspace A100 80GB instance, $92669188$ KB ($>88$ GB) general RAM.}
  \label{fig:noisy_a100}
\end{center}
\end{figure}

Figure \ref{fig:noisy_a100} shows the final fidelity estimates, for 100 random circuits at each circuit layer depth. In each trial, starting from SDRP value of $1$ and decrementing by $0.025$ at each successful completion of a circuit, we used only the fidelity estimate of the minimum attainable rounding parameter 
 $p$, before out-of-memory failure. These simulations were carried out on a single (80 GB) NVIDIA Tesla A100 GPU. Execution time was not precisely recorded, but the whole of data collection for this plot took less than or about 3 days.

The obtained results demonstrate that Qrack can run approximate simulations of 54-qubit random circuits up to 10 layers (with exponentially decreasing fidelity).  We highlight that, at 7 layers, the estimated average fidelity is  $\approx 4\%$, which is a significant result for a single GPU.

Each ``circuit layer'' is defined by a round of 3-parameter single-qubit general unitary gates,
\begin{align}\begin{aligned}\\\begin{split}U(\theta, \phi, \lambda) =
    \begin{pmatrix}
        \cos\left(\frac{\theta}{2}\right)          & -e^{i\lambda}\sin\left(\frac{\theta}{2}\right) \\
        e^{i\phi}\sin\left(\frac{\theta}{2}\right) & e^{i(\phi+\lambda)}\cos\left(\frac{\theta}{2}\right)
    \end{pmatrix}\end{split}\end{aligned}\end{align}
\cite{qiskit2017u}, with variational parameters randomly generated on their full period, acted on every qubit, followed by nearest-neighbor coupler gates from the set [CX/CY/CZ/AX/AY/AZ] (where ``A'' opposed to ``C'' indicates that the $|0\rangle$ control state activates the gate, as opposed to $|1\rangle$) applied on random qubits according to the ABCDCDAB pattern deemed to be hard to simulate in the Sycamore quantum supremacy experiment  \cite{arute2019quantum}. It should be noted that our median and mode fidelity appear significantly lower than our reported mean fidelity, to basic human inspection of our data supplement (at \url{https://github.com/unitaryfund/qrack-report}). Additional information about the fidelity obtained for different values of the SDRP approximation parameter is provided in Appendix \ref{appendix-benchmarks}.

\section{Conclusions}

We presented and numerically tested the optimization techniques which are at the basis of the Qrack simulator, many of which appear to be novel among the available set of major quantum computer simulator libraries and frameworks. We run numerical experiments for benchmarking the simulation performances on large circuits with limited computational power.

Our results show that Qrack reaches approximate parity (or better performances) relative to all other simulator candidates on exact QFT simulation, even with its hardest initial conditions (see Fig.\ \ref{fig:qft}). 
At high qubit widths, Qrack even outperforms the popular pyFFTW bindings for the (CPU-based) FFTW library, which is historically-notable for its DFT performance.
An interesting research question, suggested by our QFT benchmarks, is whether quantum-inspired (classical) algorithms for DFT could outperform standard {\it fast Fourier transform} (FFT) methods \cite{brigham1967fast, ferracin2022efficiently}, such as the {\it Cooley-Tukey} \cite{brigham1967fast} algorithm. 

For what concerns the task of approximate simulation, we gave a detailed description of the SDRP technique in which a rounding parameter $p$ can be tuned to increase the simulation efficiency at the cost of reducing the simulation fidelity. By using the SDRP technique, we have achieved $\approx 4\%$ average fidelity on random circuits acting on 54 qubits with a depth of 7 layers, a performance which is worse than Sycamore quantum supremacy experiment \cite{arute2019quantum} ($\mathcal F= 0.2\%$ with 20 layers), but remarkable considering that it was obtained with a single GPU device.

\section*{Acknowledgements}
% =============================================================================
This work was supported by the U.S. Department of Energy, Office of Science, Office of Advanced Scientific Computing Research, Accelerated Research in Quantum Computing under Award Number DE-SC0020266 as well as by IBM under Sponsored Research Agreement No. W1975810.
AM acknowledges support from the PNRR
MUR project PE0000023-NQSTI. 

\bibliographystyle{IEEEtran}
\bibliography{refs.bib}

\clearpage

\appendix
\section{Appendix}
\subsection{Empirically validating the fidelity estimation model}
\label{appendix-validation}

In this appendix we empirically validate our {\it model} of the approximate simulation fidelity $\mathcal F$ proposed in Eq.\ \eqref{eq:fid_model} against the corresponding {\it gold standard} $\mathcal F_{\rm exact}$ defined in Eq.\ \eqref{eq:fid_exact}. A data supplement which contains all validation data and additional validation tests is available at \url{https://github.com/unitaryfund/qrack-report/tree/main/data/collated}.

The main motivation for introducing and using the fidelity model $\mathcal F$ is that, when compared to $F_{\rm exact}$, it is much more efficient to compute and it becomes the only practical option for large-scale circuits as those simulated in Fig.\ \ref{fig:noisy_a100}. It is worth mentioning that an indirect approach for estimating the fidelity, although based on a  different empirical computational estimation, was also used in the quantum supremacy experiment of Ref.\ \cite{arute2019quantum} due to the prohibitive cost of computing $F_{\rm exact}$. 

For our validation we generate 100 random instances of circuits (with the structure defined in Sec.\ \ref{subsec:circuits}) for each of the following width$\times$depth combinations: 6$\times$6, 12$\times$6, 18$\times$6, 12$\times$12, 18$\times$12, 18$\times$18, 15$\times$15, and 19$\times$10. For each circuit, the Schmidt decomposition rounding parameter (SDRP) is incremented on the inclusive interval $[0, 1]$, by increments of $0.025$, interpolating from the minimum to the maximum approximation error. In total, this produced 32,800 observations of both the gold standard $\mathcal F_{\rm exact}$ and our fidelity estimation model $\mathcal F$. These circuit sizes  are chosen due to the fact that the gold standard values is possible to simulate and calculate on a single GPU.

As a validation metric we compute the root-mean-square error (RMSE) between the predictor $\mathcal F$ and the gold standard $\mathcal F_{\rm exact}$:

\begin{equation}\label{eq:rmse}
{\rm RMSE} =\sqrt{ \langle (F_j - F_{\rm exact})^2 \rangle_S } ,
\end{equation}
where $\langle  \cdot \rangle_S$ is the statistical average with respect to a set $S$ of validation circuits.

When $S$ is the set of all the validation circuits, the RMSE is $3.9\%$, in units equivalent to overall circuit fidelity. If we look at the subsets of each single width$\times$depth combination, each subset result is comparable to the overall result, as the reader can inspect in Table \ref{qrack-rmse} and in the data supplement. 

We notice that fidelities, by definition, fall on the interval $[0, 1]$. As a result, some heteroscedasticity is observed in the prediction residuals, corresponding to ceiling and floor effects on the bounded interval. When a situation like this arises, it is common to use {\it logistic regression}: the fidelity on interval $[0, 1]$ is effectively transformed with the {\it logit} function into a {\it log odds ratio} which is unbounded and suitable to the strictest assumptions of ordinary least squares regression. This alternative approach is developed and reported in the data supplement, and it points to the same essential conclusions.

\begin{table}[!htb]
\caption{\label{qrack-rmse}RMSE per width/depth subset, 100 random circuits apiece}
\begin{center}
\begin{tabular}{ c c c }
\toprule
\textbf{Width} & \textbf{Depth} & \textbf{RMSE} \\
\hline
$6$ & $6$ & $6.7\%$ \\
$12$ & $6$ & $5.4\%$ \\
$18$ & $6$ & $4.4\%$ \\
$12$ & $12$ & $3.9\%$ \\
$18$ & $12$ & $2.5\%$ \\
$18$ & $18$ & $1.8\%$ \\
$15$ & $15$ & $2.4\%$ \\
$19$ & $10$ & $2.5\%$ \\
\hline
\multicolumn{2}{c}{\textbf{Overall}} & $3.9\%$
\end{tabular}
\end{center}
\end{table}

 The data supplement, which can be found in the in the folder \url{https://github.com/unitaryfund/qrack-report/tree/main/data/collated}, is organized in different  spreadsheets named like ``validation\_method\_comparison\_n\_by\_m'', where ``n'' is circuit width and ``m'' is circuit layer depth.

\subsection{Additional simulation benchmarks}
\label{appendix-benchmarks}
This section attempts to give a more general characterization of the overall abilities of Qrack, through additional cross-sectional plots.

Figure \ref{fig:cpu-gpu} compares CPU-only methods with ``hybrid'' GPU methods in Qrack, on the quantum Fourier transform algorithm on a GHZ state. The near-identical points at the low-width end, between both series, are where hybrid GPU techniques rely on CPU, while the high-width points where GPU leads is where hybrid GPU methods actually engage the GPU.

\begin{figure}[!ht]
\begin{center}
  \includegraphics[width= 1.1 \linewidth]{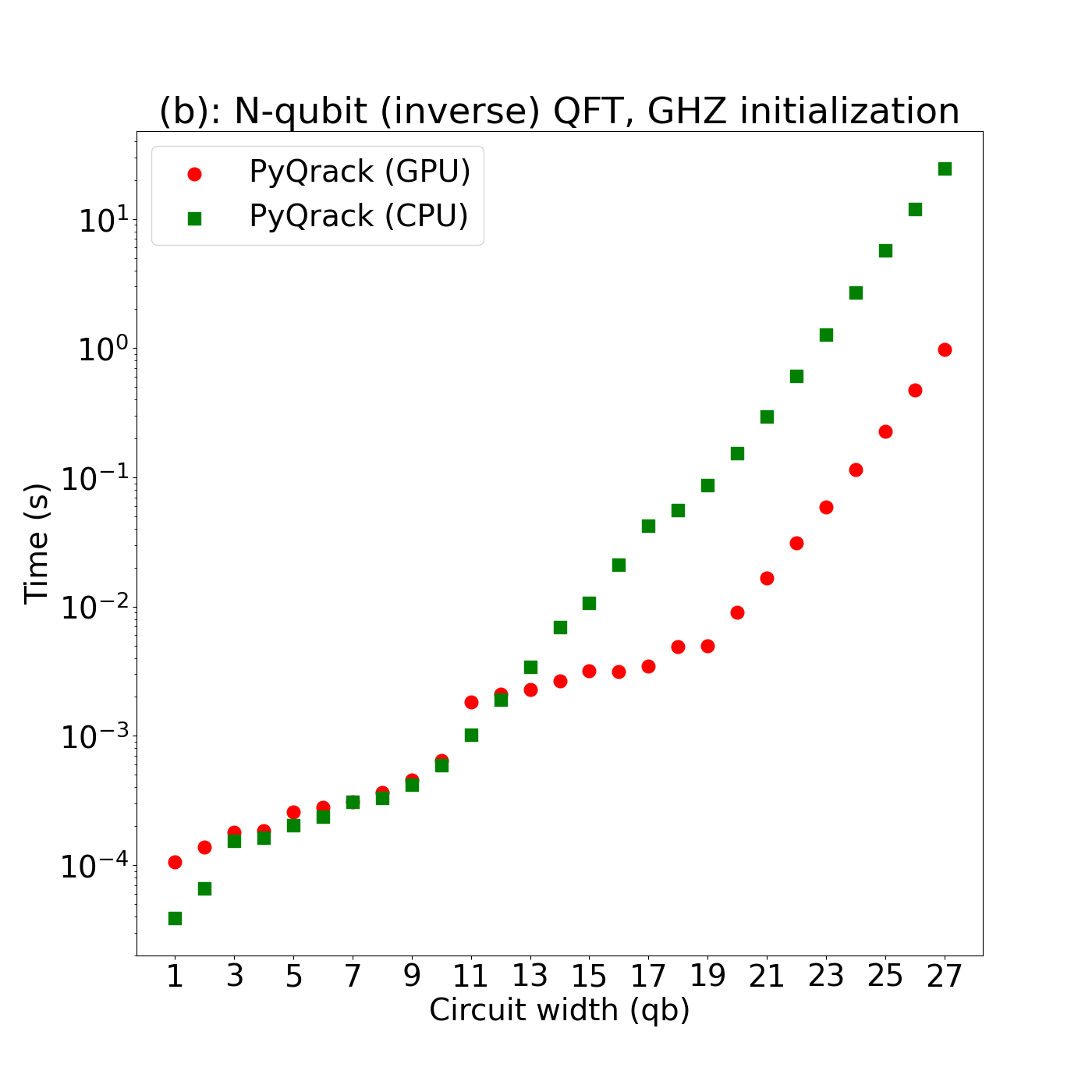}
  \caption{Comparison of Qrack executtion time with and without GPU acceleration on the quantum Fourier transform algorithm on a GHZ state. Both series were run on anAlienware m17 laptop, with Alienware BIOS version 1.16.2, (BIOS overclocking features set to off/default,) Ubuntu 22.04 LTS, Linux kernel version 5.19.0-35-generic, one ``Intel(R) Core(TM) i9-10980HK CPU @ 2.40GHz,'' one ``NVIDIA GeForce RTX 3080 Laptop GPU,'' and 32 GB of SK hynix 3200 MT/s DDR4 in 2x16 GB row configuration, with PyQrack release 1.4.2}
  \label{fig:cpu-gpu}
\end{center}
\end{figure}

\begin{figure}[!ht]
\begin{center}
  \includegraphics[width= 1.1 \linewidth]{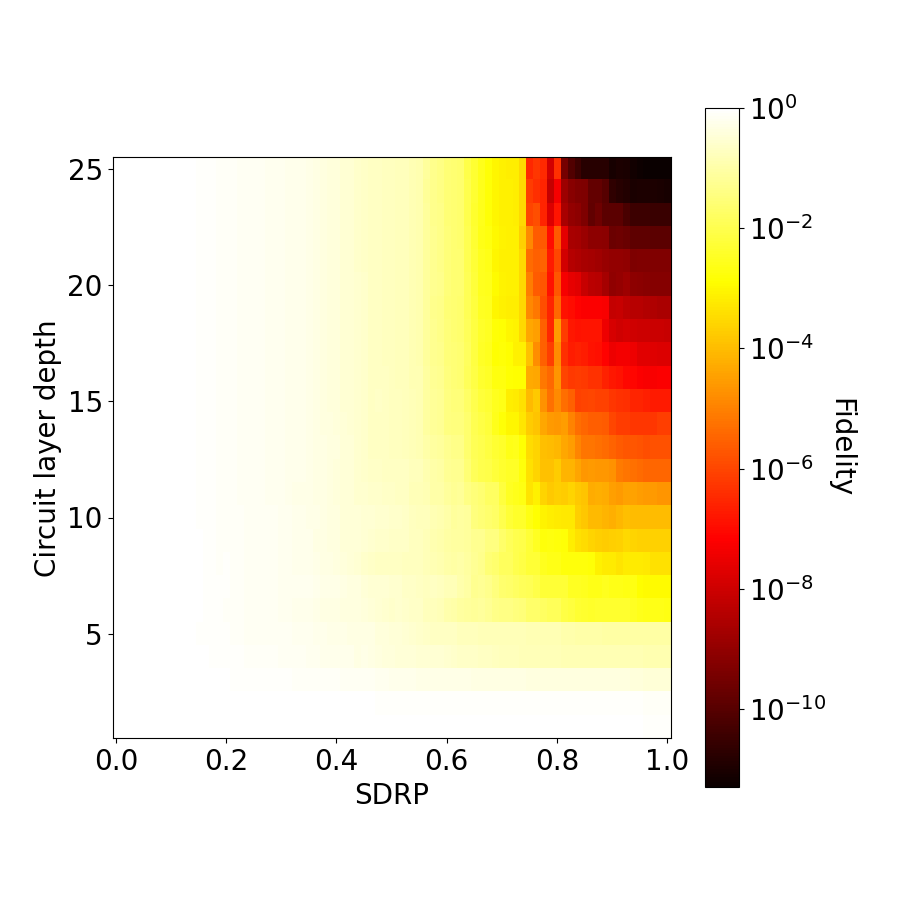}
  \caption{Random circuit fidelity on an A100 GPU, at 25 qubits width.}
  \label{fig:cpu-gpu}
\end{center}
\end{figure}

In the additional three heat map plots of SDRP, we run the same random circuit as figure \ref{fig:noisy_a100}, but rather than search for the minimum attainable rounding parameter and report the average fidelity, we give a cross section over the fidelity as we vary rounding parameter and depth. These cross sections were collected at different qubit widths: 25, 36, and 49. (Note that 49 qubits width had no significant fidelity beyond about 12 circuit layers in depth, and the black regions on the left sides of two of the heat maps ran out of memory despite potentially low SDRP values.)

\begin{figure}[!ht]
\begin{center}
  \includegraphics[width= 1.1 \linewidth]{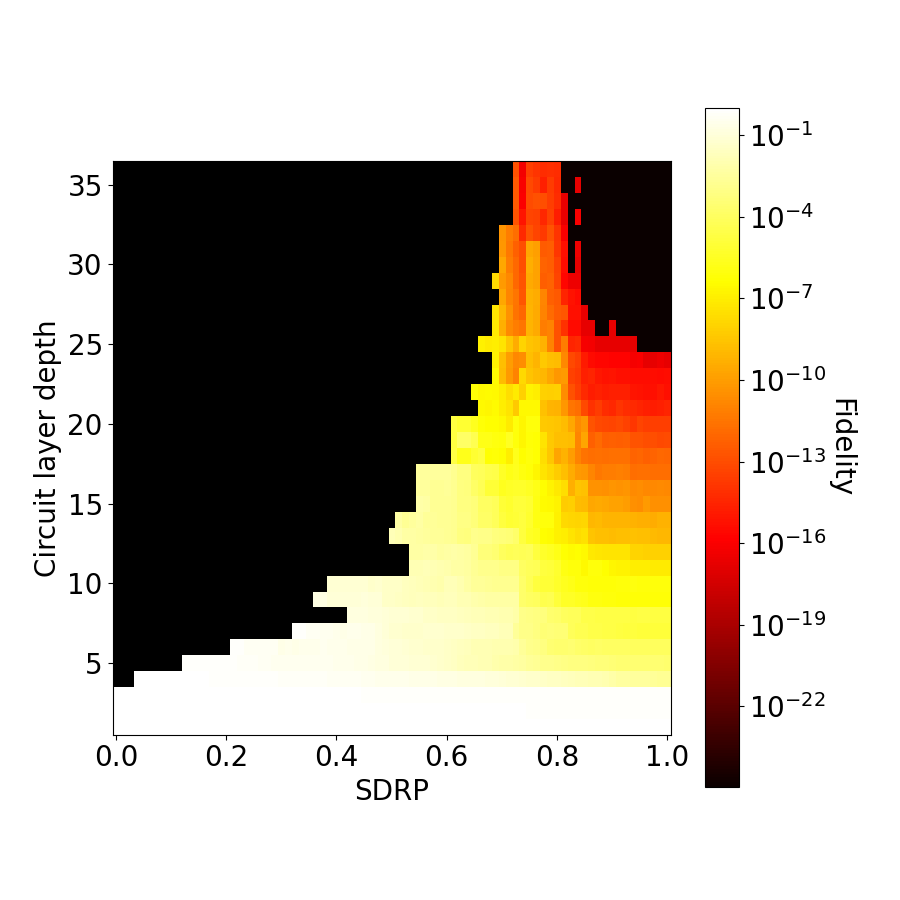}
  \caption{Random circuit fidelity on an A100 GPU, at 36 qubits width. (The black region on the left side of the heat map ran out of memory despite potentially low SDRP values.)}
  \label{fig:cpu-gpu}
\end{center}
\end{figure}

\begin{figure}[!ht]
\begin{center}
  \includegraphics[width= 1.1 \linewidth]{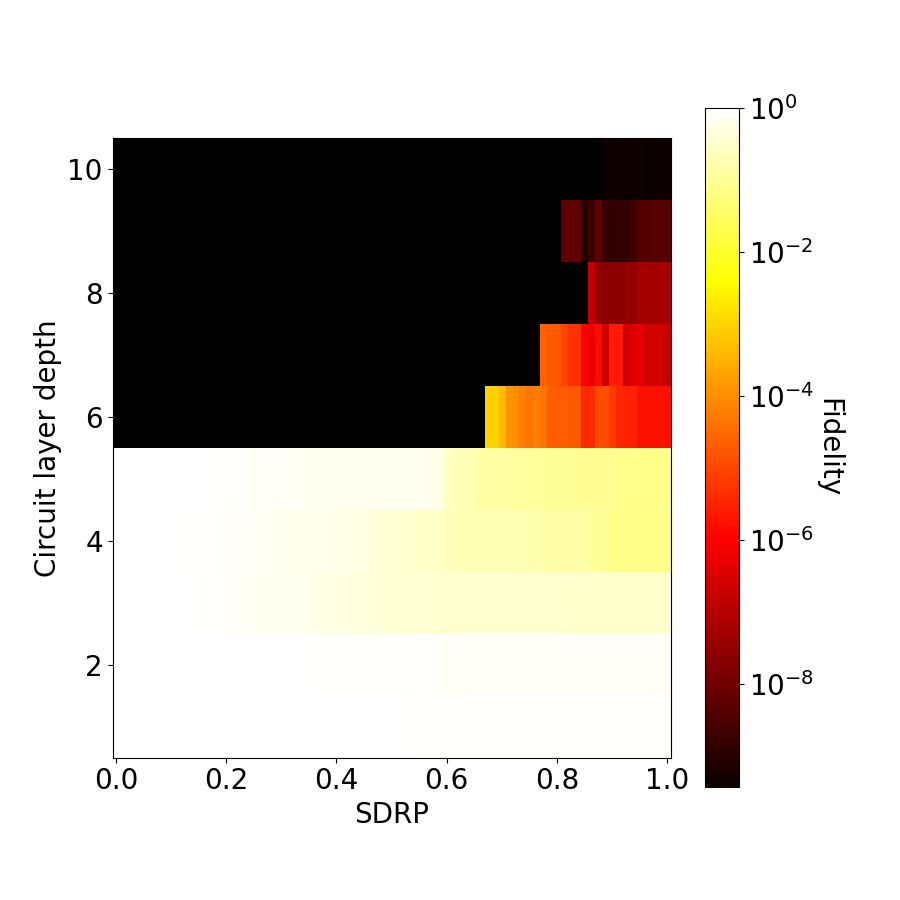}
  \caption{Random circuit fidelity on an A100 GPU, at 49 qubits width. (The black region on the left side of the heat map ran out of memory despite potentially low SDRP values.)}
  \label{fig:cpu-gpu}
\end{center}
\end{figure}

\end{document}